\documentclass[12pt]{book}

\usepackage[dvips]{graphicx,color}
\usepackage{makeidx,tsukuba}

\makeauthorindex
\makeindex

\begin{document}

\BookTitle{\itshape The 28th International Cosmic Ray Conference}
\CopyRight{\copyright 2003 by Universal Academy Press, Inc.}
\pagenumbering{arabic}

\chapter{Study of the Effect of Neutrino Oscillation on the SuperNova Neutrino Signal with the LVD Detector}

\author{%
%
%
M. Selvi$^1$ on behalf of the LVD Collaboration and F. Vissani, {\it LNGS - INFN.}
\\{\it (1) Bologna University and INFN.} 
}

\section*{Abstract}
We present an update of our previous study [2] on how $\nu$ oscillations affect the signal from a supernova core collapse observed in the LVD detector at LNGS. In this paper we use a recent, more precise determination of the cross section [8] to calculate the expected number of inverse beta decay events, we introduce in the simulation also the $\nu$-{\rm Fe} interactions, we include the Earth matter effects and, finally, we study also the inverted mass hierarchy case.

\section{Supernova neutrino signal}
At the end of its burning phase a massive star ($M \ge 8 M_\odot$) explodes into a supernova (SN), originating a neutron star which cools emitting about 99~\% of
the liberated gravitational binding energy
in neutrinos.
The time-integrated spectra can be well approximated by the
pinched Fermi--Dirac distribution,
 with an effective degeneracy parameter
$\eta$, which is assumed, for simplicity, to be null.
Although the hierarchy
$\langle E_{\nu_e}\rangle < \langle E_{\bar\nu_e}\rangle < \langle
 E_{\nu_x}\rangle $ remains valid
(here $\nu_x$ refers to both $\nu_\mu$ and $\nu_\tau$),
recent studies with an improved treatment of $\nu$ transport,
microphysics, the inclusion of the nucleon bremsstrahlung, and the
energy transfer by recoils, find somewhat smaller differences between the
$\bar\nu_e$ and $\nu_x$ spectra [5].

In the following, we assume a future galactic SN explosion at a typical
distance of $D = 10$~kpc, with a binding energy of $E_b = 3 \cdot
10^{53}$~erg and a total energy of $E^{\rm tot}_{\nu_e} = E^{\rm
tot}_{\bar\nu_e} = E^{\rm tot}_{\nu_x}$.
We also assume that the fluxes of
$\nu_\mu$, $\nu_\tau$, $\bar\nu_\mu$, and $\bar\nu_\tau$ are identical, we
fix 
$T_{\nu_x} / T_{\bar{\nu}_e} = 1.5$, 
$T_{\nu_e} / T_{\bar{\nu}_e}  = 0.8$
and $T_{\bar{\nu}_e} 
=5~{\rm MeV}$ [5].

\section{LVD detector and neutrino reactions}

\begin{figure}[t]
  \begin{minipage}{.48\columnwidth}
    \begin{center}
      \includegraphics[height=17pc]{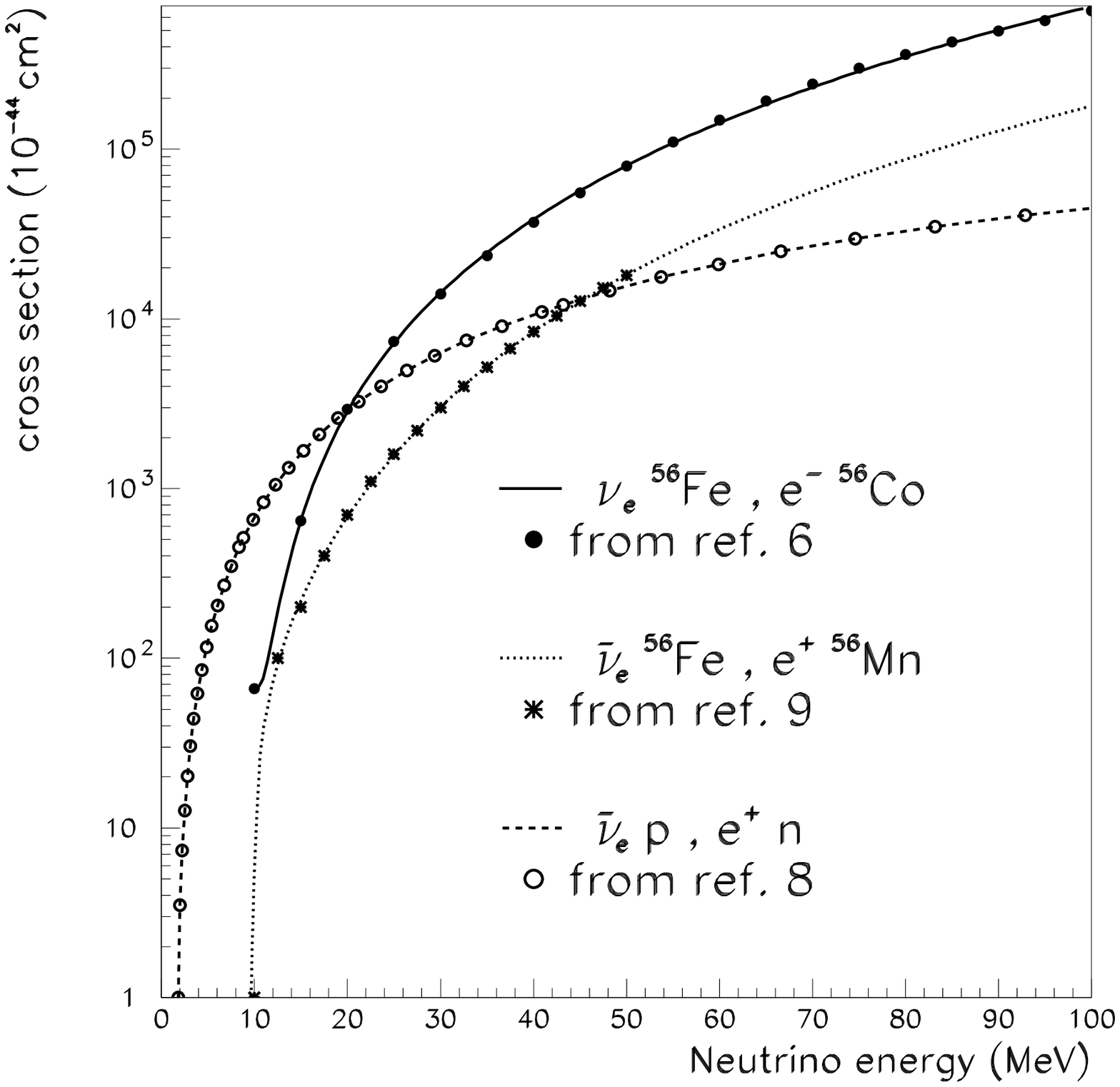}
    \end{center}
    \vspace{-1.cm}
    \caption{Cross sections of the main neutrino interactions in LVD.}
    \label{fi:sig}
  \end{minipage}
  \hspace{1pc} 
  \begin{minipage}{.48\columnwidth}
    \begin{center}
      \includegraphics[height=17pc]{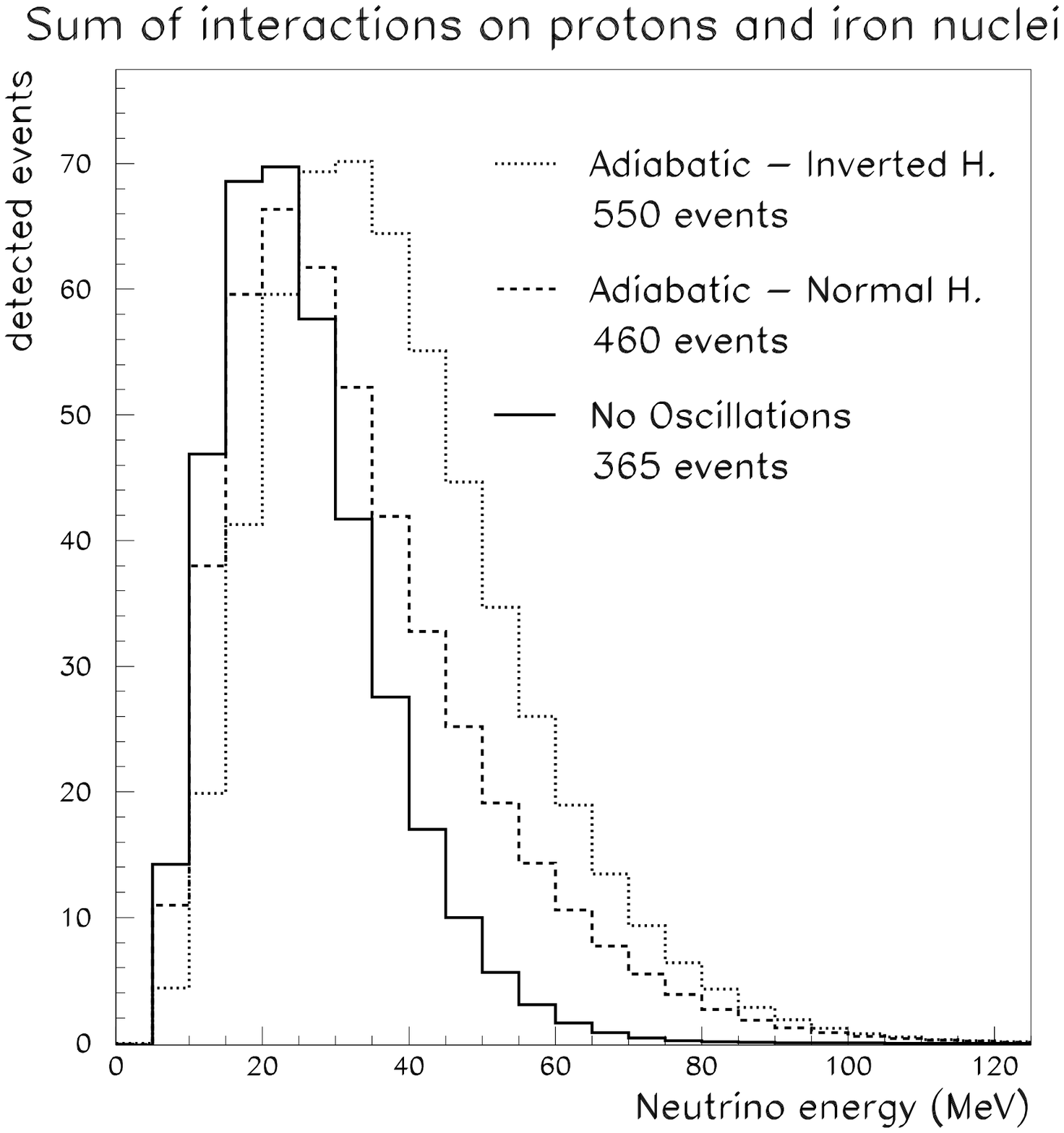}
    \end{center}
    \vspace{-1.cm}
    \caption{Effect of neutrino oscillations in the signal detected in LVD.}
    \label{fi:osc}
  \end{minipage}
\end{figure}

The Large Volume Detector (LVD) in the INFN Gran Sasso National Laboratory,
Italy, consists of an array of 840  
liquid scintillator (LS) counters, 1.5 m$^3$ each. These are 
interleaved by streamer tubes, and arranged in a 
modular 
geometry; a detailed description is in [1].
The active scintillator mass is $M=1000$ t. 
There are two subsets of counters: the
external ones ($43 \%$),
operated at energy threshold ${\cal E}_h\simeq 7$ {\rm MeV},
and inner ones ($57 \%$), better shielded from rock radioactivity and 
operated at ${\cal E}_h\simeq 4$ {\rm MeV}. 
In order to tag the delayed $\gamma$ pulse due to
$n$-capture, all counters are equipped with an additional discrimination
channel, set at a lower
threshold, ${\cal E}_l\simeq 1$ {\rm MeV}.

In the following we will focus on $\nu$ reactions with free protons and iron nuclei:
 {\em (1)}\
$\bar\nu_e p,e^+ n$, observed through a prompt signal from ${e}^+$ above
threshold ${\cal E}_h$ (detectable energy $E_d \simeq E_{\bar\nu_{e}}-1.8$ {\rm MeV}
$+ 2 m_e c^2 $), followed by the signal from the ${n p,d} \gamma$ capture
($E_{\gamma} = 2.2$ {\rm MeV}), above ${\cal E}_l$ and with a mean delay
$\Delta t \simeq 180~\mu \mathrm{s}$.  The cross section for this reaction has been recently recalculated [8] with a better treatment of the $10-100~{\rm MeV}$ region, i.e. the SN neutrino energy.
The cross section behaviour with energy is shown in figure \ref{fi:sig} The efficiency for the prompt signal is $\epsilon_{\bar\nu_e\, p,e^+\, n} = 95\% $. The total number of free protons in the LS is $9.34~\cdot 10^{31}$.

The LVD detector presents an iron support structure made basically by two components: the tank (mean thickness: $0.4 ~ cm$) which contains the LS and the portatank (mean thickness: $1.5 ~ cm$) which hosts a cluster of 8 tanks. Indeed, the higher energy part of the $\nu$ flux could be detected also with the $\nu (\bar\nu)  {\rm Fe}$ interaction, which results in an electron (positron) that could exit iron and release energy in the LS. The considered  reactions are:\\
\noindent {\em (2)}\ $\nu_e\,^{56}\mathrm{Fe},^{56}\!\mathrm{Co}\ e^-$. The binding energy difference between the ground levels is $E_b^{\rm{Co}} - E_b^{\rm{Fe}} = 4.566~{\rm MeV}$; moreover the first $\rm{Co}$ allowed state is at $3.589~{\rm MeV}$. Indeed, in this work we considered  $E_{e^-} = E_{\nu_e} - 8.15~ {\rm MeV}$. 
A full simulation of the LVD support structure and LS geometry has been developed in order to get the efficiency for an electron, generated randomly in the iron structure, to reach the LS with energy higher than ${\cal E}_h$. It is greater than $20\%$ for $E_\nu > 30~{\rm MeV}$ and grows up to $70\%$ for $E_\nu > 100~{\rm MeV}$ . On average, the electron energy detectable in LS is $E_d \simeq 0.45 \times E_\nu$. The total number of iron nuclei is $7.63~\cdot10^{30}$.

\noindent {\em (3)}\ $\bar\nu_e\,^{56}\mathrm{Fe},^{56}\!\mathrm{Mn}\ e^+$,
the energy threshold is very similar to reaction {\em (2)} and the same considerations could be done.
The cross section for reactions {\em (2),(3)} are taken respectively from [6,9] and plotted in figure \ref{fi:sig}



\section{Neutrino oscillation and MSW effect in the SN and in the Earth}
In the study of SN neutrinos,
$\nu_{\mu}$ and $\nu_{\tau}$ are indistinguishable, both in the star and in
the detector because of the corresponding charged lepton production threshold; consequently, in the frame of three-flavor oscillations,
the relevant parameters are just
$(\Delta m^2_{{\rm sol}}, U_{e2}^2)$ and 
$(\Delta m^{2}_{\rm atm}, U_{e3}^2)$. 
We will adopt the following numerical values:
$\Delta m^2_{{\rm sol}}=7 \cdot 10^{-5}{\rm eV}^2$, 
$\Delta m^{2}_{\rm atm}=2.5 \cdot 10^{-3} {\rm eV}^2$, 
$U_{e2}^2=0.33;$ the selected solar parameters 
$(\Delta m^2_{{\rm sol}}, U_{e2}^2)$ describe the
LMA-I solution, as it results from a global analysis including solar, CHOOZ and KamLAND $\nu$ data [4].

For a normal mass hierarchy (NH) scheme,  
$\nu$ (not $\bar \nu$) cross two resonance layers:
one at higher density (H), which corresponds to $\Delta m^{2}_{\rm atm}, U_{e3}^2$, and
the other at lower density (L), corresponding to 
$\Delta m^{2}_{{\rm sol}}, U_{ e2}^2$.
For inverted mass hierarchy (IH), transitions at
the higher density layer occur in the $\bar \nu$
sector, while at the lower density layer they occur 
in the $\nu$ sector. 

\begin{figure}[t]
  \begin{minipage}{.48\columnwidth}
    \begin{center}
      \includegraphics[height=17pc]{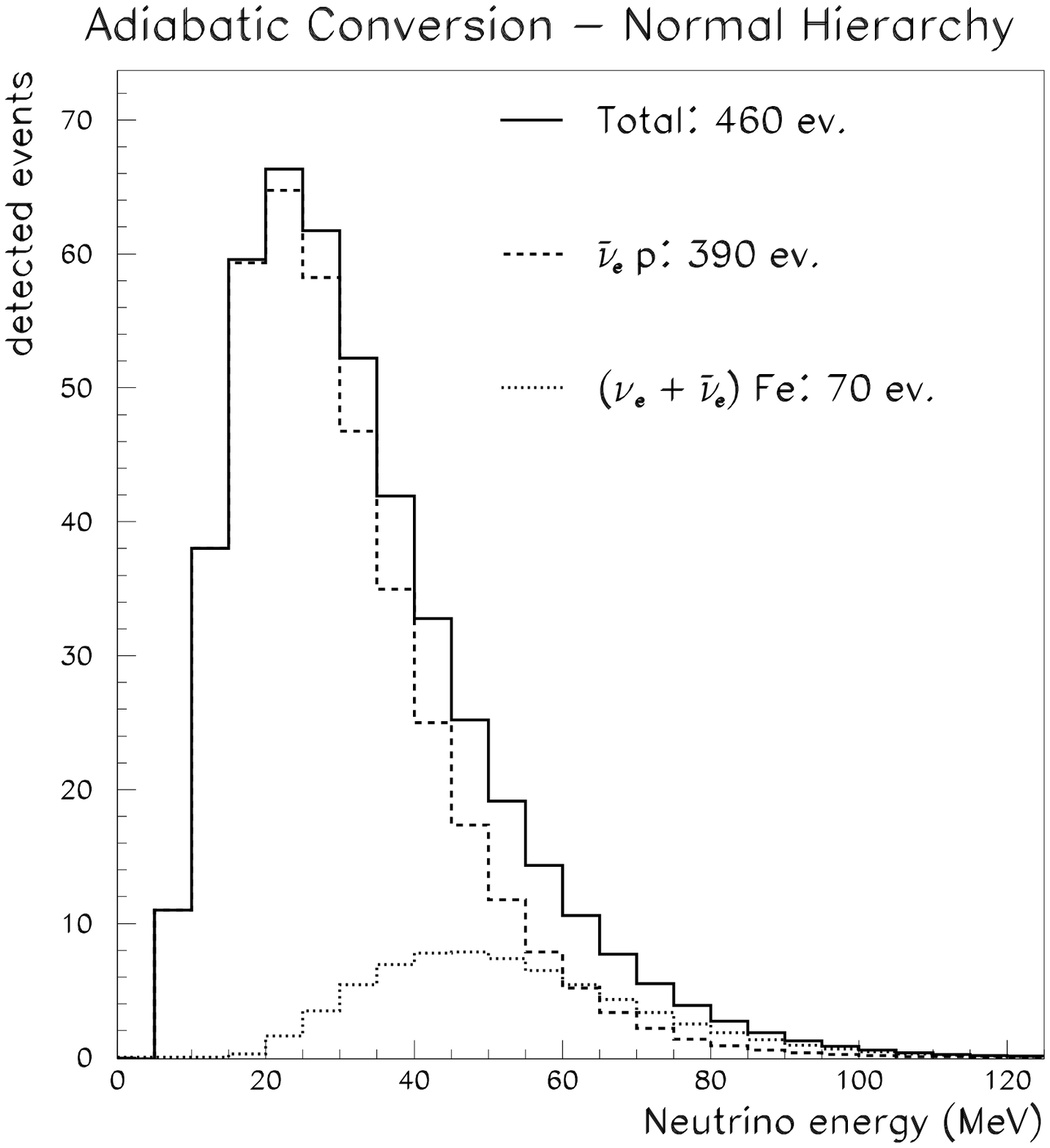}
    \end{center}
    \vspace{-1.cm}
    \caption{Impact of iron interactions in the global neutrino signal in LVD.}
    \label{fi:iron}
  \end{minipage}
  \hspace{1pc} 
  \begin{minipage}{.48\columnwidth}
    \begin{center}
      \includegraphics[height=17pc]{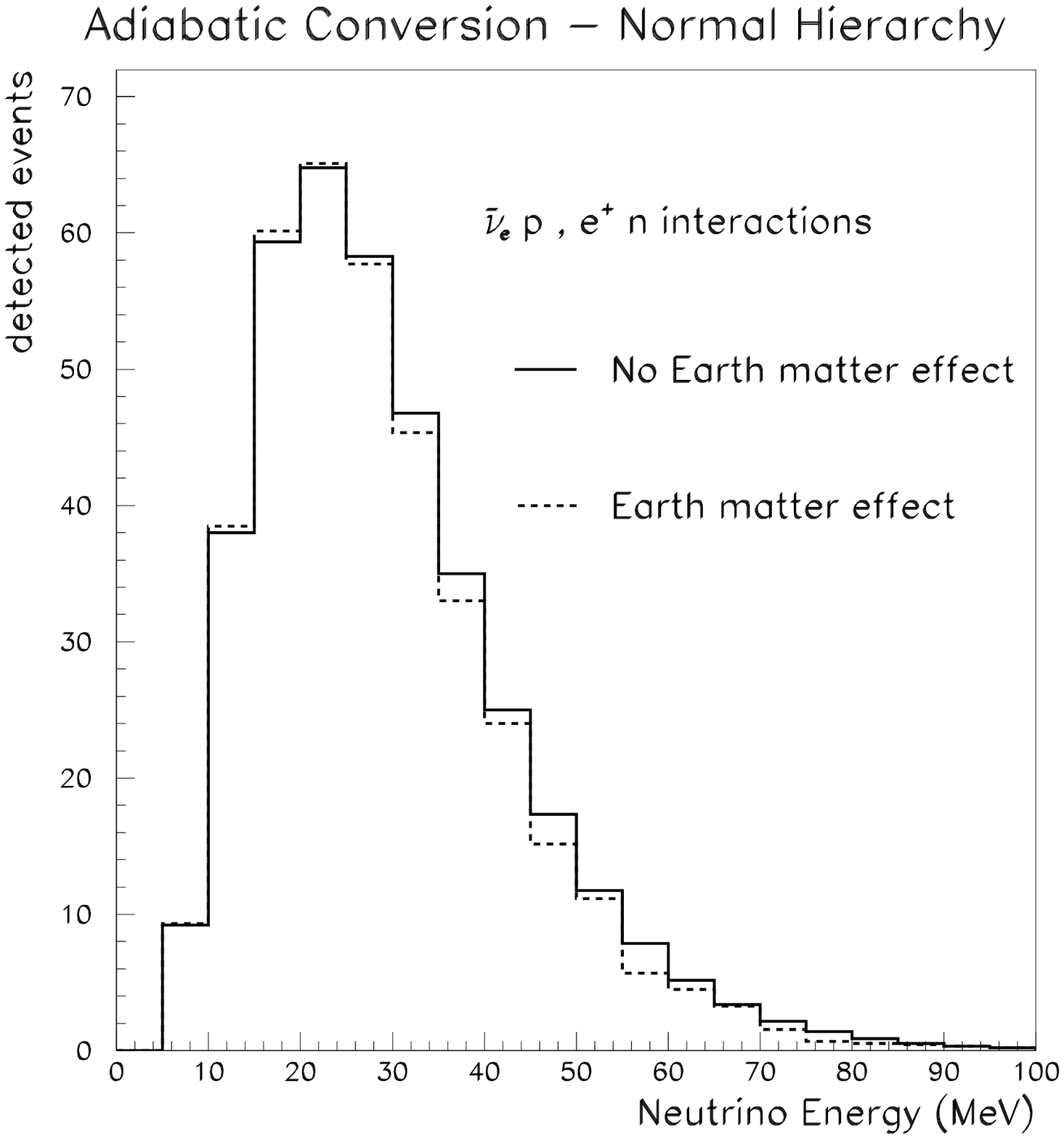}
    \end{center}
    \vspace{-1.cm}
    \caption{Effect of the Earth matter in the $\bar\nu_e p,e^+ n$ signal in LVD.}
    \label{fi:mswe}
  \end{minipage}
\end{figure}

Given the energy range of SN $\nu$ (up to $\sim 100~{\rm MeV}$)
and considering a star density profile $\rho \propto 1/r^3$, 
the adiabaticity condition is always satisfied at the L resonance
for any LMA solution, while at the H resonance, this 
depends on the value of $U_{e3}^2$. 
When $U_{e3}^2 \geq 5 \cdot 10^{-4}$ 
the conversion is completely adiabatic, meaning that the flip probability between two adiacent mass eigenstates is null ($P_h=0$).
In the adiabatic case and NH, the $\bar \nu_e$ produced in the SN core arrive at Earth as $\nu_1$, and they have a high ($U_{e1}^2 \simeq cos^2 \theta_{12} \simeq 0.7$) probability to be detected as $\bar\nu_e$. 
On the other hand, the original $\bar \nu_x$ arrive at Earth as $\nu_2$ and $\nu_3$ and are detected as $\bar \nu_e$ with probability $U_{e2}^2 \simeq sin^2 \theta_{12}$. Given the higher energy spectrum of $\bar \nu_x$ this configuration results in a larger number of interactions, with respect to the no-oscillation case, due to the increasing cross sections with energy. In the adiabatic-IH case the detected $\bar \nu_e$ completely come from the original $\bar \nu_x$ flux in the star and the number of interaction is still greater, as shown in figure \ref{fi:osc}
The oscillations scheme can be summarized as:
 $F_e = P_h U_{e2}^2 F_e^0 + (1-P_h U_{e2}^2) F_x^0$ and  $F_{\bar e} = U_{e1}^2 F_{\bar e}^0 + U_{e2}^2 F_{\bar x}^0$ for normal hierarchy; 
$F_e = U_{e2}^2 F_e^0 + U_{e1}^2 F_x^0$ and $F_{\bar e} = P_h U_{e1}^2 F_{\bar e}^0 + (1 - P_h U_{e1}^2) F_{\bar x}^0$ for inverted hierarchy, 
where $F_{any}^0$ are the original neutrino fluxes in the star and $F_{any}$ are the observed $\nu$ fluxes. One can notice that, in the antineutrino channel, the non adiabatic ($P_h=1$), IH case, is equivalent to the NH case (which does not depend on adiabaticity). 
In figure \ref{fi:iron} is shown the contribution of $(\nu_e+\bar \nu_e)$ {\rm Fe} interactions in the total number of events. For the chosen SN and oscillation parameters they are about $18\%$ of the signal.

If we consider the effect of Earth in the neutrino path to the detector, we must replace, in the detected flux estimation, $U_{ei}^2$ with $P_{ei}~ (i=1,2)$, the probability for the mass eigenstate $\nu_i$ to be detected as $\nu_e$ after path in the Earth [7], which depends on the solar oscillation parameters and on the travelled density profile through the Earth.
We developed a complete 3-flavour calculation, describing the earth interior as made of 12 equal density steps, following the PREM matter density profile. For each constant density step we compute the exact propagator of the evolution matrix and we get the global amplitude matrix by multiplying the propagators of the traversed density layers, following the strategy of [3].
In figure \ref{fi:mswe} the effect of Earth matter in the SN neutrino signal is shown for a nadir angle $\theta_n=50^\circ$, which corresponds to neutrinos passing through the mantle only. Earth matter effect are more relevant in the $\nu$ than in the $\bar \nu$ channel, so the effect in reaction {\em (1)} is quite weak (it also depends on the rather high $\Delta m^2_{{\rm sol}}$), but it could be detected if compared with a high statistic sample (i.e. SuperKamiokande) or if a larger number of events is available, i.e. a closer SN.

%




\section{References}




\re
1. M. Aglietta et al., {\em Il Nuovo Cimento A} {\bf 105} (1992), 1793.

\re
2. M. Aglietta et al., {\em Nucl. Phys. B Proc. Suppl.} {\bf 110} (2002), 410-413.

\re
3. E. Kh. Akhmedov, {{\tt hep-ph/0001264}}.


\re
4. G. L. Fogli et al.,  {{\tt hep-ph/0212127}}.

\re
5. M.~T. Keil, G.~G. Raffelt, and H.-T. Janka, {{\tt astro-ph/0208035}}.

\re
6. E. Kolbe, K. Langanke, {{\tt nucl-th/0003060}}.

\re
7. C. Lunardini, A. Yu. Smirnov,  {{\tt hep-ph/0106149}}.




\re
8. A. Strumia, F. Vissani, {{\tt astro-ph/0302055}}.


\re
9. J. Toivanen et al., {\em Nuclear Physics A} {\bf 694} (2001), 395-408.


\endofpaper
\end{document}